\documentclass[%
 reprint,
 nofootinbib,
 nobibnotes,
 bibnotes,
 amsmath,amssymb,
prb,
]{revtex4-2}

\usepackage{color}%
\usepackage{graphicx}
\usepackage{dcolumn}
\usepackage{bm}
\usepackage{hyperref}
\usepackage[mathlines]{lineno}

\begin{document}


\title{Charging in the vortex lattice of type-II superconductors}

\author{Marie Ohuchi,$^1$ Hikaru Ueki,$^{1,2}$ and Takafumi Kita$^1$}%
\affiliation{%
$^1$Department of Physics, Hokkaido University, Sapporo, Hokkaido 060-0810, Japan\\
$^2$Department of Mathematics and Physics, Hirosaki University, Hirosaki, Aomori 036-8561, Japan
}%

\date{\today}

\begin{abstract}
We study the magnetic-field dependence of the vortex-core charge in the Abrikosov lattice of an $s$-wave superconductor based on the augmented quasiclassical equations, where we incorporate the pair-potential gradient (PPG) terms characteristic of charging in superconductors besides the well-known Lorentz force. Our numerical results at $T=0.2 T_{\rm c}$ and $0.5 T_{\rm c}$ reveal that periodic charge redistribution is superimposed on the magnetic flux-line lattice with different spatial patterns at different fields. The PPG terms are dominant at weak fields over the Lorentz force for accumulating charge in the vortex cores, whereas the Lorentz force prevails at higher fields to give rise to a peak structure in the core charge around $H\sim \frac{1}{2}H_{{\rm c}2}$. We estimate the peak value of the core charge at $T=0.2 T_{\rm c}$ using parameters appropriate for cuprates to obtain a large value of $Q \sim 10^{-2} |e|$ in the core region of radius $0.2 \xi_0$ in the $ab$ plane and length $1 \ {\rm nm}$ along the $c$ axis. 
\end{abstract}

\maketitle


\section{Introduction \label{sec:I}}
It is well known that vortices in type-II superconductors 
have the magnetic character of carrying a single flux quantum per each of them.
In contrast, much less familiar may be the fact that they also have an electric feature 
with accumulation of charge
in the core region due to circulating supercurrents and pair-potential reduction around it. 
The earliest studies on the vortex-core charging were carried out 
based on phenomenological approaches \cite{Khomskii95,Blatter96}, 
which were followed by the microscopic ones of using
the Bogoliubov--de Gennes (BdG) equations \cite{Hayashi98,MH,Machida03}. 
London included the Lorentz force acting on supercurrents 
in his phenomenological equations of superconductivity,
which predict vortex-core charging
due to the Lorentz force \cite{LondonText,Kita09}.  
Khomskii and Freimuth estimated the vortex-core charge phenomenologically 
by regarding the core region as the normal state distinguishable by a radial step function 
and considering its
chemical potential difference from the outer superconducting region \cite{Khomskii92,Khomskii95}. 
Matsumoto and Heeb pioneered a microscopic calculation based on the BdG equations
coupled with Maxwell's equations to predict vortex-core charging 
in an isolated vortex of a chiral $p$-wave superconductor \cite{MH}. 
The phenomenon was also studied
by using the Ginzburg--Landau (GL) Lagrangian 
that additionally incorporates the Chern--Simons term \cite{Goryo}. 
On the other hand, Eschrig {\it et al}.\ \cite{Eschrig99,Eschrig09} calculated a dynamical dipole charge 
in the vortex core 
under an applied AC electromagnetic field. 
It was also shown that electric charge accumulates even at the vortex core 
of electrically neutral $p$-wave superfluids,
although the magnitude is much smaller than the one around a core of superconductors \cite{Volovik}.
Experimentally, the vortex-core charge in cuprate superconductors was estimated 
using nuclear magnetic resonance/nuclear quadrupole resonance (NMR/NQR) measurements \cite{Kumagai01}.

The BdG approach to charging in superconductors has a firm microscopic basis 
but also suffers from a shortcoming of being time-consuming and laborious numerically.
Thus, quantitative studies of charging in superconductors remain yet to be performed, 
especially for vortex-lattice states in magnetic fields. To this end, 
augmented quasiclassical (AQC) equations 
of superconductivity with quantum corrections 
were derived recently by collecting next-to-leading-order 
contributions in the expansion of the Gor'kov equations \cite{Gor'kov59,Gor'kov60} 
in terms of the quasiclassical parameter $\delta\equiv1/k_{\rm F}\xi_0$ \cite{UOK18}, where 
$k_{\rm F}$ and $\xi_0$ are the Fermi wavenumber and zero temperature coherence length, 
respectively. 
This quasiclassical approach has elucidated three distinct mechanisms for charging in superconductors: 
(i) the Lorentz force that acts on supercurrents in magnetic fields \cite{Kita01,UKT};
(ii) pair-potential gradient (PPG) terms \cite{AK14,OUK17,UOK18};
and (iii) terms originating from the slope in the density of states (SDOS) 
\cite{Khomskii92,Khomskii95,UOK18}, the latter two of which are characteristic of superconductors.
The resulting AQC equations were used to clarify charging of an isolated vortex 
in $s$-wave superconductors
with cylindrical  \cite{OUK17} and spherical \cite{UOK18} Fermi surfaces.
The outcome was rather surprising in that it is the PPG terms, not the Lorentz force, 
that contributes mainly to charging of an isolated vortex core,
except near $T_{\rm c}$ with a large GL parameter $\kappa_{\rm GL}$ 
where the SDOS terms become dominant \cite{UOK18}. 
Masaki also studied charging of an isolated vortex in $s$- and chiral $p$-wave superconductors
based on the AQC equations with the Lorentz force and PPG terms \cite{Masaki18}
to obtain results consistent with those based on the BdG equations \cite{MH}. 
On the other hand, the AQC equations with {\it only the Lorentz force} were used 
for vortex-lattice states of $s$-wave \cite{KUK16} 
and  $d$-wave \cite{KUK17} superconductors to study the magnetic-field dependence of charging. 
It was thereby shown that the charge density at the core has a large peak  
as a function of the magnetic field. 
These results naturally raise the question of how the field dependence is affected 
by including the PPG terms in the AQC equations,
i.e., an issue that should be answered in a quantitative manner.

The purposes of the present paper are twofold: (i) to develop a numerical method 
for calculating charging 
in the Abrikosov lattice \cite{Abrikosov57} of type-II superconductors 
within the AQC scheme incorporating both the Lorentz force and PPG terms;
and (ii) to elucidate magnetic-field dependence of the vortex-core charge. 
To avoid numerical complexity as far as possible, we adopt the simplest model of $s$-wave pairing on 
a cylindrical Fermi  surface with the field applied along the cylinder, 
where the SDOS terms does not contribute to charging at all
due to the constant density of states \cite{OUK17}. 
We combine the methods developed 
in Refs.\ \cite{OUK17} and \cite{KUK16} to perform numerical calculations 
of charging in the Abrikosov lattice.

This paper is organized as follows. 
In Sec.\ \ref{sec.II}, we present our formalism 
to study charging in superconductors based on the AQC equations. 
In Sec.\ \ref{sec.III}, we give numerical results 
on charging in the vortex lattice 
to clarify field dependence and relative magnitude of charging due to the Lorentz force and PPG terms. 
Section \ref{sec.IV} gives a conclusion.

{\section{Augmented quasiclassical equations\label{sec.II}}}
We consider a clean superconductor with $s$-wave pairing on 
a cylindrical Fermi surface in a magnetic field applied along the cylinder, omitting spin paramagnetism for simplicity.
The corresponding AQC equations in the Matsubara formalism with the Lorentz force and PPG terms 
 are given by \cite{OUK17,UOK18}
{\setlength\arraycolsep{2pt} 
\begin{eqnarray} 
\bigl[ i \varepsilon_n \hat{\tau}_3 &-& \hat{\Delta}\hat\tau_3, \hat{g} \bigr] + i  \hbar {\bf v}_{\rm F} \cdot {\bm\partial} \hat{g} \notag \\ 
&+& \frac{i  \hbar}{2}e ({\bf v}_{\rm F} \times {\bf B}) \cdot \frac{\partial}{\partial {\bf p}_{\rm F}} \bigl\{ \hat{\tau}_3, \hat{g} \bigr\} \notag \\
&-&\frac{i\hbar}{2}{\bm\partial}\hat{\Delta}\hat{\tau}_3\cdot\frac{\partial \hat{g}}{\partial {\bf p}_{\rm F}} -\frac{i\hbar}{2}\frac{\partial \hat{g}}{\partial {\bf p}_{\rm F}}\cdot{\bm\partial}\hat{\Delta}\hat{\tau}_3 = \hat{0}. \label{LandP} 
\end{eqnarray}}%
Here $\hat{g}\!=\!\hat{g}(\varepsilon_n,{\bf p}_{\rm F},{\bf r})$ and $\hat{\Delta}\!=\!\hat\Delta({\bf  r})$ are the quasiclassical Green's functions and the pair potential, respectively; $\varepsilon_n\!=\!(2n+1)\pi k_{\rm B}T$ is the fermion Matsubara energy $(n=0,\pm 1,\cdots)$ with $k_{{\rm B}}$ and $T$ denoting the Boltzmann constant and temperature; ${\bf v}_{\rm F}$ and ${\bf p}_{\rm F}$ are the Fermi velocity and momentum; $e < 0$ is the electron charge; ${\bf B}={\bf B}({\bf r})$ is the magnetic-flux density; the commutators are defined by $[ \hat{a}, \hat{b} ] \equiv \hat{a} \hat{b} - \hat{b} \hat{a}$ and $\{ \hat{a}, \hat{b} \} \equiv \hat{a} \hat{b} + \hat{b} \hat{a}$; and ${\bm \partial} $ is the gauge-invariant differential operator,
\begin{align} 
{\bm\partial} \equiv \left\{ \begin{array}{ll} 
\vspace{2mm}{\bm\nabla} &  {\rm on} \  g \ {\rm or} \ \bar{g}, \\
\displaystyle \vspace{2mm}{\bm\nabla} - i \frac{2 e {\bf A}}{\hbar} &  {\rm on} \ f \ {\rm or} \ \Delta, \\
\displaystyle  {\bm\nabla} + i \frac{2 e {\bf A}}{\hbar} & {\rm on} \ \bar{f} \ {\rm or} \ \Delta^*,
\end{array}\right. 
\end{align} 
with ${\bf A}\!=\!{\bf A}({\bf r})$ denoting the vector potential. 
The first line in Eq.\ (\ref{LandP}) forms 
the standard Eilenberger equations \cite{Eilenberger,KitaText,SR,LO86},
the second line represents the Lorentz force \cite{Kita01,UKT}, 
and the third line is the PPG terms \cite{AK14,OUK17,UOK18}.  
The matrices $\hat{g}$, $\hat{\Delta}$, and $\hat{\tau}_3$ are expressible as \cite{KitaText}
\begin{align}
\hat{g}  =
\begin{bmatrix}
\vspace{1mm}
g & - if \\
i\bar{f} & - \bar{g}
\end{bmatrix}, \hspace{3mm}
\hat{\Delta} =
\begin{bmatrix}
\vspace{1mm}
0 & \Delta \\
\Delta^{\!*} & 0
\end{bmatrix},\hspace{3mm}
\hat{\tau}_3 =
\begin{bmatrix}
\vspace{1mm}
1 & 0 \\
0 & -1
\end{bmatrix},
\end{align}
where the barred functions are defined generally by $\bar{X} (\varepsilon_n, {\bf p}_{\rm F}, {\bf r}) \equiv X^* (\varepsilon_n, - {\bf p}_{\rm F}, {\bf r})$.

Following the procedure used in Ref.\ \cite{Kita09}, 
we expand $g$ and $f$ formally in terms of $\delta\equiv1/k_{\rm F}\xi_0$ 
as $g=g_0+g_1+\cdots$ and $f=f_0+f_1+\cdots$, 
where $g_0$ and $f_0$ are the solutions of the standard Eilenberger equations,
and $\xi_0$ is defined by $\xi_0\equiv\hbar v_{\rm F}/\Delta_0$ in terms of the energy gap $\Delta_0$ at zero magnetic field and zero temperature. 
Collecting zeroth-order terms in Eq.\ (\ref{LandP}) reproduces the standard Eilenberger equations with the normalization condition 
$g_0\!=\!{{\rm sgn}}(\varepsilon_n)\bigl(1-f_0\bar{f}_0\bigr)^{1/2}$ 
as \cite{Eilenberger,KitaText,SR,LO86,KopninText} 
\begin{subequations}
\label{Ei}
\begin{align} 
&\varepsilon_n f_0+\frac{1}{2}\hbar{\bf v}_{{\rm F}}\cdot \left({\bm\nabla} - i \frac{2 e {\bf A}}{\hbar}\right) f_0=\Delta g_0, \label{Ei1} \\
&\Delta=\Gamma_0 \pi  k_{{\rm B}} T \sum_{n=-\infty}^{\infty}\langle f_0 \rangle_{{\rm F}} \label{Ei2}, \\
&{\bm \nabla}\times{\bm \nabla}\times{\bf A}=\mu_0{\bf j}, 
\notag \\ 
& {\bf j}=-i2\pi eN(0)k_{{\rm B}}T\sum_{n=-\infty}^{\infty}\langle {\bf v}_{{\rm F}}g_0 \rangle_{{\rm F}} \label{Ei3}, 
\end{align}
\end{subequations}
where ${\bf j}={\bf j}({\bf r})$ is the current density, 
$\Gamma_0\ll 1$ is the dimensionless coupling constant responsible for the Cooper pairing, 
$\langle\cdots\rangle_{\rm F}$ 
is the Fermi-surface average normalized as $\langle 1\rangle_{\rm F}=1$, 
$\mu_0$ is the vacuum permeability, 
and $N(0)$ is the normal density of states (DOS) per spin 
and unit volume at the Fermi energy. 
Equation (\ref{Ei}) forms a set of self-consistent equations for $f_0$, $\Delta$, and ${\bf A}$.

The equation for $g_1$ can be obtained from Eq.\ (\ref{LandP}) as \cite{OUK17,UOK18} 
\begin{align} 
{\bf v}_{\rm F} \cdot {\bm\nabla} g_1 =& - e ({\bf v}_{\rm F} \times {\bf B}) \cdot \frac{\partial g_0}{\partial {\bf p}_{\rm F}} \notag \\
&- \frac{i}{2} {\bm\partial}\Delta^{\!*} \cdot  \frac{\partial f_0}{\partial {\bf p}_{\rm F}} - \frac{i}{2} {\bm\partial} \Delta \cdot  \frac{\partial \bar f_0}{\partial {\bf p}_{\rm F}}, \label{nablag1}
\end{align}
with $g_1 = -\bar{g}_1$. 
The electric field ${\bf E} ={\bf E}({\bf r})$ obeys \cite{UOK18} 
\begin{align} 
&- \lambda_{\rm TF}^2 {\bm\nabla}^2 {\bf E} + {\bf E} = i  \frac{\pi k_{\rm B} T}{e} \sum_{n = - \infty}^\infty \left\langle {\bm\nabla}g_1 \right\rangle_{\rm F} \notag \\
& \ \ \ -\frac{1}{e} \frac{N'(0)}{N(0)}\int_{-\tilde{\varepsilon}_{\rm c}}^{\tilde{\varepsilon}_{\rm c}}d\varepsilon \bar{n} (\varepsilon){\varepsilon}\left\langle{\bm\nabla}{\rm Re}g_0^{\rm R} \right\rangle_{\rm F}
{-\frac{c}{e}}\frac{N'(0)}{N(0)}{\bm\nabla} {|\Delta|^2}, \label{EEq}
\end{align}
where $\lambda_{\rm TF} \!\equiv\! \sqrt{\epsilon_0/2 e^2 N (0)}$ 
is the Thomas--Fermi screening length with $\epsilon_0$ denoting 
the vacuum permittivity, and the function 
$\bar{n} (\varepsilon)={1}/{({\rm e}^{\varepsilon/k_{\rm B}T}+1})$ 
is the Fermi distribution function. 
The first term on the right-hand side of Eq.\ (\ref{EEq}) represents charging by the Lorentz force and PPG terms, while the second and third terms are contributions from the SDOS terms. 
The constant $c$ is the factor introduced by Khomskii and Freimuth\ \cite{Khomskii95,UOK18},
\begin{align}
c\equiv\int_{-\tilde{\varepsilon}_{\rm c}}^{\tilde{\varepsilon}_{\rm c}}d\varepsilon\frac{1}{2\varepsilon}\tanh\frac{\varepsilon}{2k_{\rm{B}}T_{\rm{c}}},
\end{align}
 where $T_{\rm{c}}$ is the superconducting transition temperature at zero magnetic field.
The cutoff energy $\tilde{\varepsilon}_{\rm c}$ is determined by \cite{UOK18} 
\begin{align}
\int_{-\tilde{\varepsilon}_{\rm c}}^{\tilde{\varepsilon}_{\rm c}}N_{\rm s}(\varepsilon,{\bf r})d\varepsilon
=\int_{-\tilde{\varepsilon}_{\rm c}}^{\tilde{\varepsilon}_{\rm c}}N(\varepsilon)d\varepsilon,
 \ \ \ N_{\rm s}(\pm\tilde{\varepsilon}_{\rm c},{\bf r}) = N (\pm\tilde{\varepsilon}_{\rm c}), 
\label{cutoffenergyeq}
\end{align}
where $N_{\rm s}(\varepsilon, {\bf r})$ and $N(\varepsilon)$ 
are the superconducting local and normal DOS, respectively. 
These equations enable us to calculate the electric field and charge density microscopically. 
As already mentioned, the SDOS terms can be dropped in the present study 
with a cylindrical Fermi surface \cite{OUK17}.

\begin{figure*}
\begin{center}
\includegraphics[width=\linewidth]{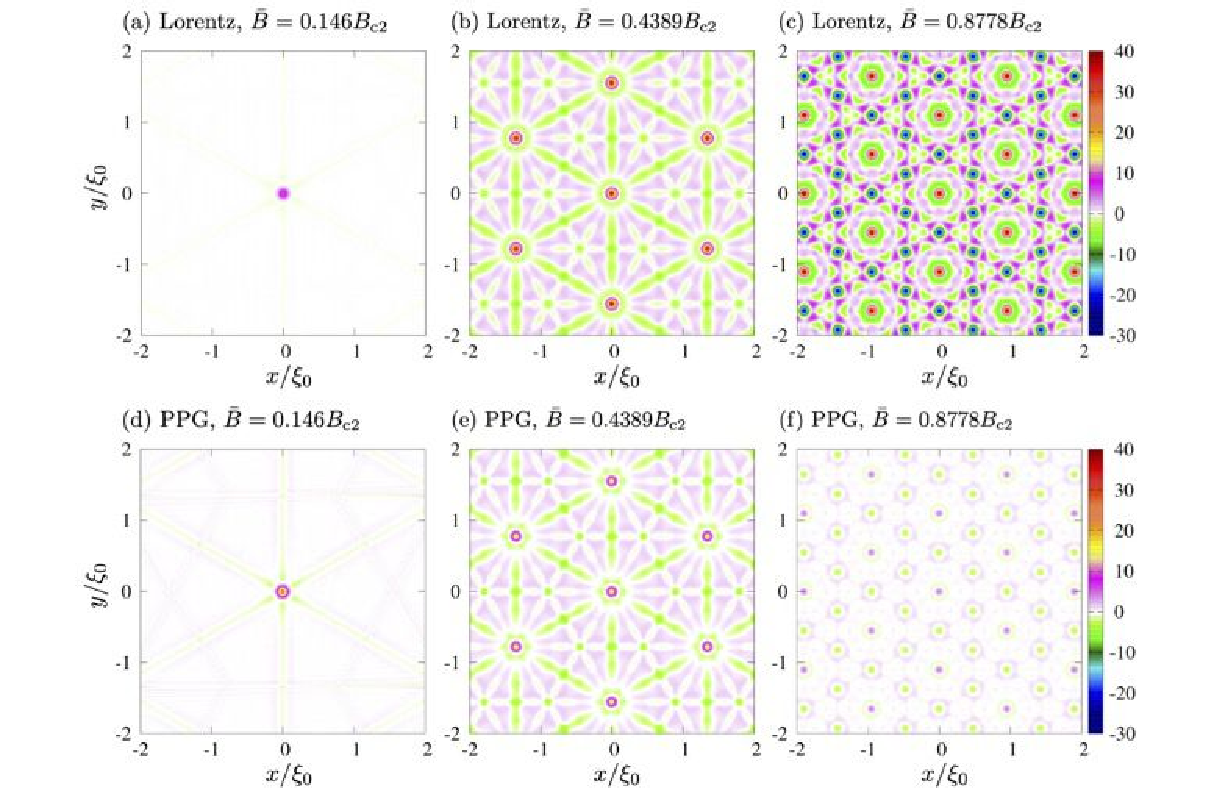}
\end{center}
\caption{Charge density $\rho({\bf r})$ due to (a)--(c) the Lorentz force 
and (d)--(f) the PPG terms 
at temperature $T=0.2T_{\rm c}$
in units of $\rho_0\equiv\Delta_0\epsilon_0/|e|\xi_0^2$ 
on a square region of  $x,y\in [-2\xi_0, +2\xi_0]$
for the average flux densities
$\bar{B} = 0.146B_{{\rm c}2}$, $0.4389B_{{\rm c}2}$, 
and $0.8778B_{{\rm c}2}$ from left to right, respectively.}
\label{fig-rhor}
\end{figure*}
\begin{figure*}
\begin{center}
\includegraphics[width=\linewidth]{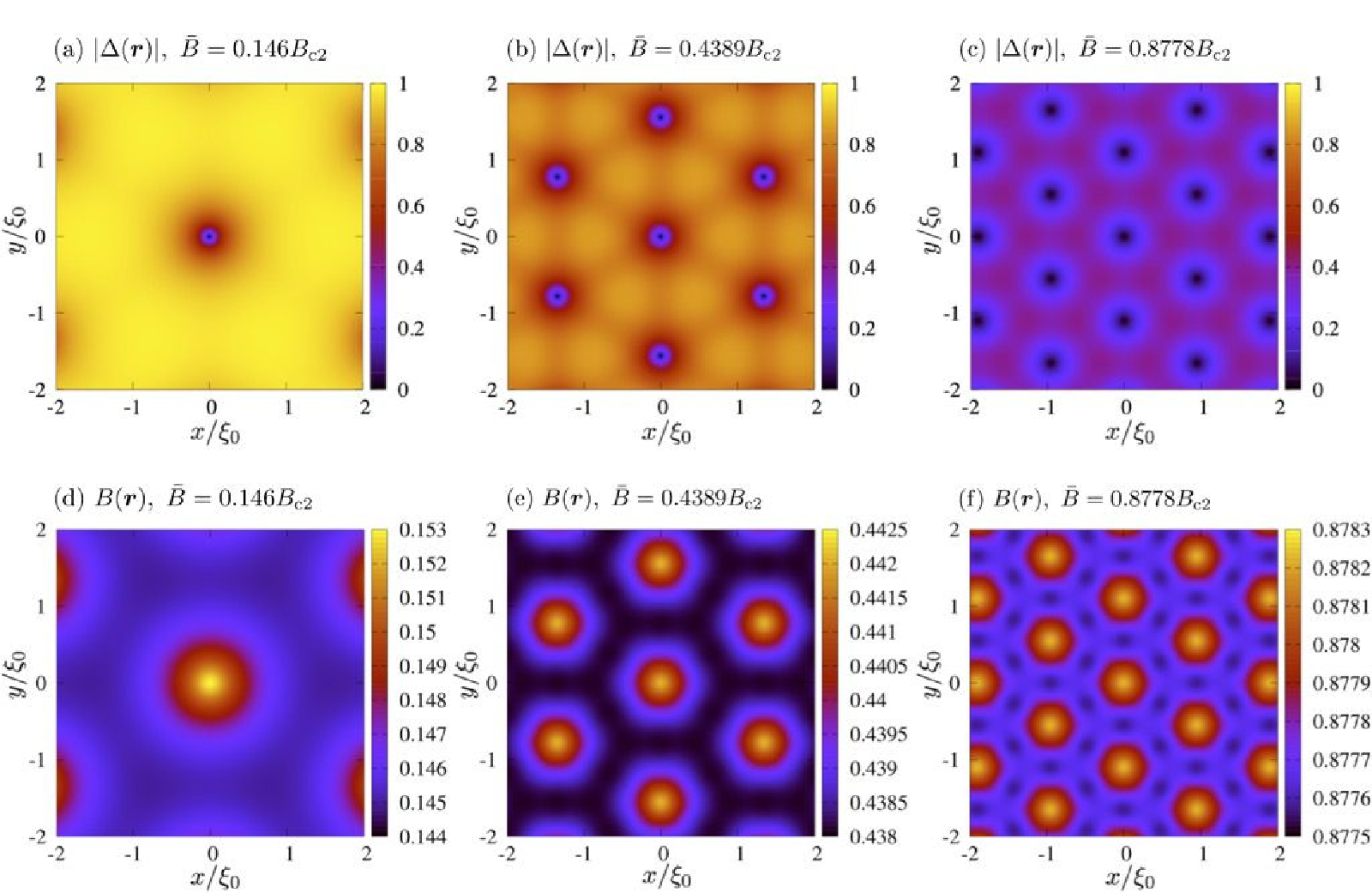}
\end{center}
\caption{(a)--(c) Gap amplitude $|\Delta({\bf r})|$  
and (d)--(f) the magnetic-flux density $B({\bf r})$ 
at temperature $T=0.2T_{\rm c}$ 
in units of the zero temperature gap $\Delta_0$ 
and the upper critical field $B_{{\rm c}2}$ 
on a square region of $x,y\in [-2\xi_0, +2\xi_0]$
for the average flux densities 
$\bar{B} = 0.146B_{{\rm c}2}$, $0.4389B_{{\rm c}2}$, 
and $0.8778B_{{\rm c}2}$ from left to right, respectively. }
\label{fig-B_Del}
\end{figure*}
\begin{figure*}[t]
\begin{center}
\includegraphics[width=\linewidth]{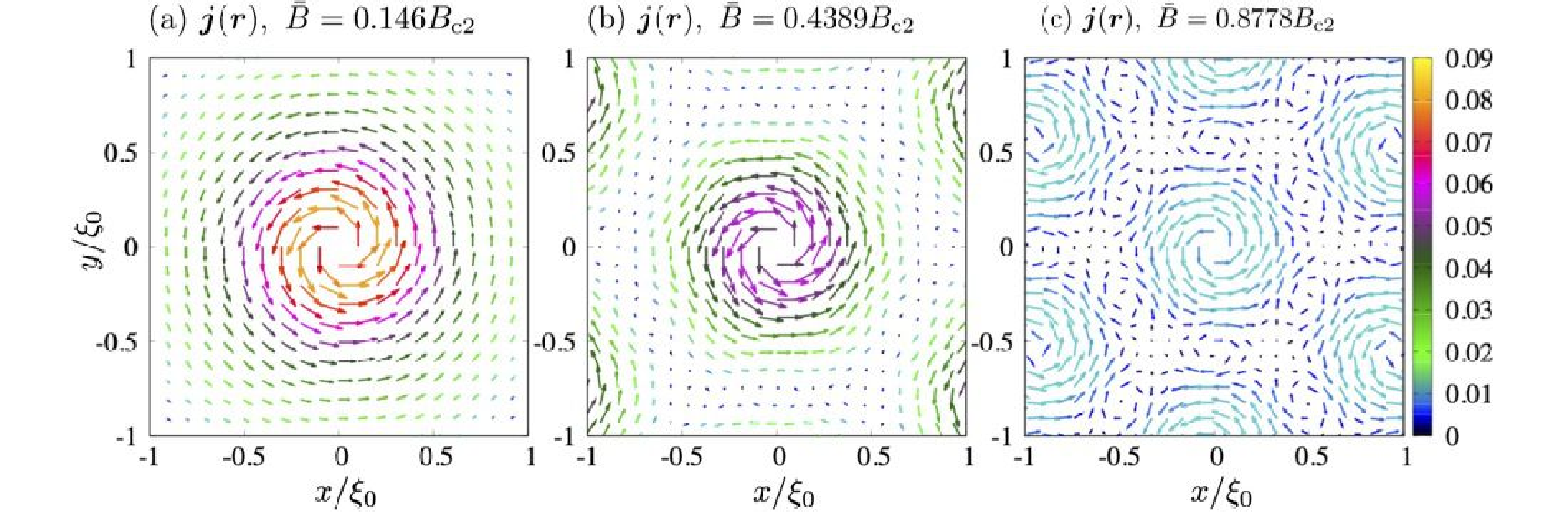}
\end{center}
\caption{Current density ${\bf j}({\bf r})$ at $T=0.2T_{\rm c}$ 
for the average flux densities
(a) $\bar{B} = 0.146B_{{\rm c}2}$, (b) $0.4389B_{{\rm c}2}$, 
and (c) $0.8778B_{{\rm c}2}$
in units of $j_0\equiv \hbar/2\mu_0|e|\xi_0^3$  
on a square region of $x,y\in [-\xi_0, +\xi_0]$. 
The color bar indicates the magnitude of the current density.}
\label{fig-j}
\end{figure*}
\begin{figure*}[t]
\begin{center}
\includegraphics[width=0.9\linewidth]{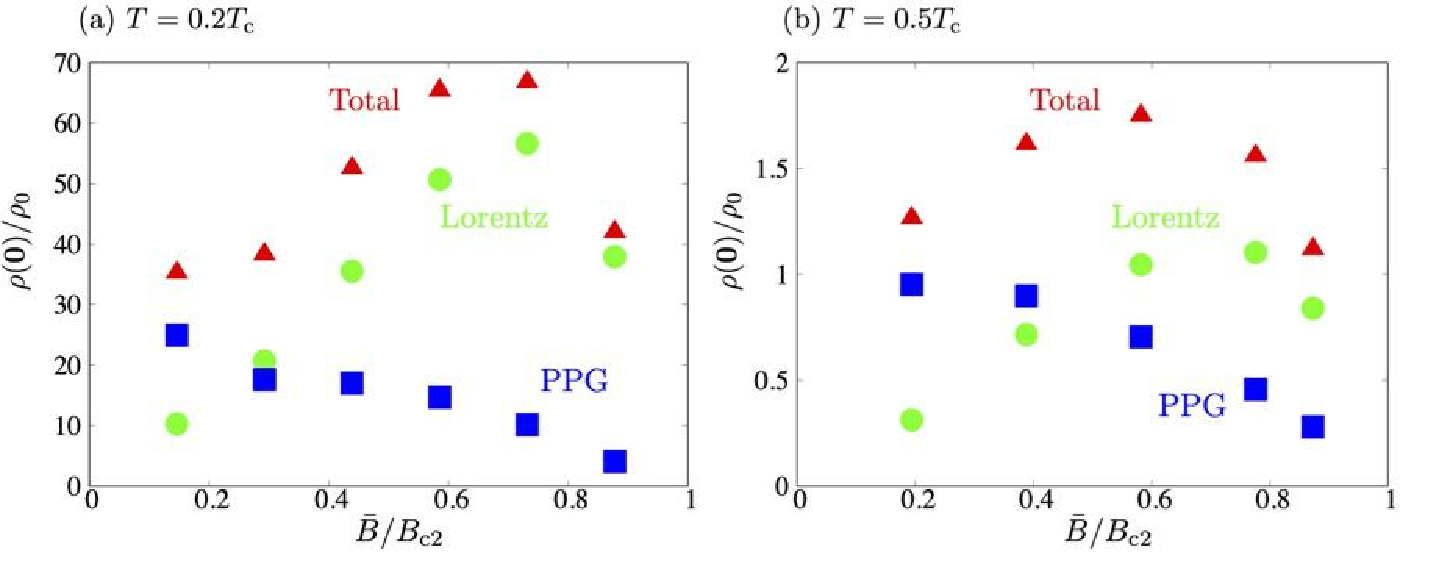}
\end{center}
\caption{
Charge density at the vortex center $\rho({\bf 0})$ 
due to the Lorentz force (green circular points), 
the PPG terms (blue square points), 
and their total (red triangular points), 
in units of  $\rho_0\equiv\Delta_0\epsilon_0/|e|\xi_0^2$ 
as a function of the magnetic field, 
at temperatures (a) $T=0.2T_{\rm c}$ and (b) $0.5T_{\rm c}$. }
\label{fig-rho0}
\end{figure*}
%

\ \
\section{Numerical Results\label{sec.III}} 
\subsection{Numerical procedures}
We solve Eqs.\ (\ref{Ei}), (\ref{nablag1}), and (\ref{EEq}) numerically 
for a triangular vortex lattice 
based on the methods in Refs.\ \cite{OUK17} and \cite{KUK16}. 
The corresponding vector potential is expressible in terms of the average flux density 
$\bar{{\bf B}}=(0,0,\bar{B})$ 
as ${{\bf A}}({\bf r})=(\bar{{\bf B}}\times{\bf r})/2+\tilde{{\bf A}}({\bf r}) $, 
where $\tilde{{\bf A}}$ describes spatial variation of the flux density that averages out. 
Functions $\tilde{\bf A}({\bf r})$, ${\bf E}({\bf r})$, and $\Delta({\bf r})$ 
obey the following periodic boundary conditions \cite{Klein,Kita98,Ichioka97}: 
\begin{subequations}
\begin{align}
&\tilde{\bf A}({\bf r}+{\bf R})=\tilde{\bf A}({\bf r}), \\
&{\bf E}({\bf r}+{\bf R})={\bf E}({\bf r}), \\
&\Delta({\bf r}+{\bf R}) 
=\Delta({\bf r}) {\rm e}^{i\chi({\bf r})}, 
\end{align}
\end{subequations}
with
\begin{align}
\chi({\bf r})\equiv&\,\frac{|e|}{\hbar}\bar{{\bf B}}\cdot({\bf r}\times{\bf R})
+\frac{|e|}{\hbar}\bar{{\bf B}}\times({\bf a}_1-{\bf a}_2)\cdot{\bf R}
+\pi n_1n_2,
\end{align}
where ${\bf R}$ is the translation vector of the vortex lattice given by 
${\bf R}=n_1{\bf a}_1+n_2{\bf a}_2$ in terms of integers $n_1$ and $n_2$, 
and ${\bf a}_1 = a_2(\sqrt{3}/2,1/2,0)$ and ${\bf a}_2 = a_2(0,1,0)$ 
are the basic vectors of the triangular lattice with the length $a_2$ 
determined by the flux-quantization condition $({\bf a}_1\times{\bf a}_2)\cdot\bar{\bf B}=h/2|e|$. 
The quasiclassical Green's functions also satisfy the periodic boundary conditions,
\begin{subequations}
\begin{align}
&g_0(\varepsilon_n, {\bf p}_{\rm F}, {\bf r}+{\bf R})=g_0(\varepsilon_n, {\bf p}_{\rm F}, {\bf r}), \\
&g_1(\varepsilon_n, {\bf p}_{\rm F}, {\bf r}+{\bf R})=g_1(\varepsilon_n, {\bf p}_{\rm F}, {\bf r}), \\
&f_0(\varepsilon_n, {\bf p}_{\rm F}, {\bf r}+{\bf R}) 
=f_0(\varepsilon_n, {\bf p}_{\rm F}, {\bf r}) {\rm e}^{i\chi({\bf r})}.
\end{align}
\end{subequations}

We first solve the standard Eilenberger equations (\ref{Ei}) self-consistently 
for the vortex lattice using the Riccati method \cite{KitaText,Nagato93,SM,Schopohl}.
The solution is substituted into the right-hand side of Eq.\ (\ref{nablag1}), 
which is solved by the Runge--Kutta method. 
We next obtain the electric field 
by solving Eq.\ (\ref{EEq}) in terms of the solution of Eq.\ (\ref{nablag1}),  
and then calculate the charge density $\rho$ 
using Gauss' law $\rho=\epsilon_0 {\bm\nabla} \cdot {\bf E}$. 
The results presented below are 
for $\lambda_{\rm TF} = 0.03 \xi_0$, $\lambda_0=5\xi_0$, and $\delta = 0.03$ 
(i.e., $\lambda_{\rm TF}\!=\! k_{\rm F}^{-1}$), 
where $\lambda_0$ is the magnetic penetration depth at zero temperature
defined by 
$\lambda_{0}\!\equiv\! \left[\mu_0N(0)e^2v_{\rm F}^2\right]^{-1/2}$. 
The magnetic flux density is normalized by the upper critical field $B_{{\rm c}2}=\mu_0H_{{\rm c}2}$
estimated by the Helfand--Werthamer theory \cite{Helfand,Kita04}.

\subsection{Results}
Figure \ref{fig-rhor} plots 
spatial dependence of the charge density $\rho({\bf r})$ due to the Lorentz force 
and PPG terms at temperature $T=0.2T_{\rm c}$ 
for the average flux densities 
$\bar{B} = 0.146B_{{\rm c}2}$, $0.4389B_{{\rm c}2}$, and $0.8778B_{{\rm c}2}$, 
respectively. 
For reference, we also give the corresponding
pair potential $\Delta({\bf r})$ and flux density $B({\bf r})$ in Fig.\ \ref{fig-B_Del}, 
which perfectly reproduce the preceding results by Ichioka {\it et al}.\ \cite{Ichioka97}.
Specifically, Fig.\ \ref{fig-B_Del} plots 
spatial variation of the gap amplitude $|\Delta({\bf r})|$ 
and the $z$-component of the magnetic-flux density $B({\bf r})$, 
respectively, and Fig.\ \ref{fig-j} shows the current density ${\bf j}({\bf r})$, each calculated
at $T=0.2T_{\rm c}$ 
for $\bar{B} = 0.146B_{{\rm c}2}$, $0.4389B_{{\rm c}2}$, and $0.8778B_{{\rm c}2}$. 
Looking at Fig.\ \ref{fig-rhor} in comparison with Figs.\ \ref{fig-B_Del} and \ref{fig-j}, 
we realize that the large and positive charges are accumulated at the vortex cores 
where the pair potential vanishes. 
We also observe that the Lorentz force becomes dominant from intermediate to high fields for accumulating charge around vortex cores, which may naturally be expected from its expression ${\bf F}_{\rm L}\propto{\bf j}\times{\bf B}$; see
Figs.\ \ref{fig-B_Del} and \ref{fig-j} on this point.
We also find in Fig.\ \ref{fig-rhor} that the negative charges are accumulated 
at the midpoint of each vortex triangle and also along the line connecting adjacent vortices. The negative charges around the midpoint can be explained in terms of the Lorentz force by looking closely at the spatial profile of the current density in Fig.\ 3(c), where the current is seen to circulate oppositely in direction to that around the vortex core.
This argument also applies to the negative charge accumulation along the lines connecting adjacent vortices in Figs.\ \ref{fig-rhor}(b) and \ref{fig-rhor}(c). 
Thus, charge accumulation due to the Lorentz force can be understood based on the force picture when the current density, magnetic field, and signs of the carriers are specified. 
On the other hand, it is difficult to explain how the force due to the PPG terms acts on electrons. This may be because the PPG terms are complex and have off-diagonal components in the particle-hole space. For clarity, we use ``the PPG terms" and not ``the PPG force" in this paper. 
Whether a force picture is also possible for the PPG terms remains a future issue, which may be solved by calculating macroscopic forces, as has been done recently based on the time-dependent GL equation \cite{Kato16,Sugai21}.

Figure \ref{fig-rho0} 
shows the magnetic-field dependence of the charge density at the vortex center 
due to the Lorentz force and PPG terms 
at $T=0.2T_{\rm c}$ and $T=0.5T_{\rm c}$, respectively. 
We can confirm that the vortex-core charge due to the Lorentz force 
has a large peak in qualitative agreement with the previous work \cite{KUK16}. 
However,  its magnitude obtained here is about 10 times larger than that given in the previous work \cite{KUK16}. This is because the previous work inappropriately neglected the component in ${\bm\nabla} g_1$ perpendicular to the Fermi velocity. Specifically, the AQC equations with only the Lorentz force yields the following equation for $g_1$,
\begin{align}
{\bf v}_{\rm F}\cdot{\bm\nabla}g_1 = -e{\bf v}_{\rm F}\cdot\left({\bf B}\times\frac{\partial g_0}{\partial {\bf p}_{\rm F}}\right). \label{g1_Lorentz1}
\end{align}
This equation was approximated previously by \cite{KUK16}
\begin{align}
{\bm\nabla}g_1 = -e\left({\bf B}\times\frac{\partial g_0}{\partial {\bf p}_{\rm F}}\right). \label{g1_Lorentz2}
\end{align}
However, ${\bm\nabla}g_1$ may have a component perpendicular to ${\bf v}_{\rm F}$. 
Hence, we directly solved the equation for $g_1$ obtained in Refs.\ \cite{OUK17,UOK18}, 
i.e., Eq.\ (\ref{nablag1}) above.

We can observe in Fig.\ \ref{fig-rho0} that the charge due to the PPG terms 
at the core center decreases monotonically as the field is increased. 
We also notice an enhancement of the vortex-core charge due to the PPG terms at $T=0.2T_{\rm c}$ and $\bar{B}=0.146B_{{\bf c}2}$
compared with those at higher fields and temperatures, which may be explained as follows.
The PPG terms in Eq.\ (\ref{nablag1}) can be written as 
\begin{align} 
&-\frac{i}{2}{\bm\nabla}|\Delta | \cdot \frac{\partial \tilde{f}_0}{\partial {\bf p}_{\rm F}} 
-\frac{i}{2} {\bm\nabla}|\Delta | \cdot \frac{\partial \bar{\tilde{f}}_0}{\partial {\bf p}_{\rm F}} \notag \\
&-\frac{m}{\hbar}|\Delta |{\bf v}_{\rm s} \cdot \frac{\partial \tilde{f}_0}{\partial {\bf p}_{\rm F}} 
+\frac{m}{\hbar} |\Delta |{\bf v}_{\rm s} \cdot \frac{\partial \bar{\tilde{f}}_0}{\partial {\bf p}_{\rm F}}, \label{nablag1PPG} 
\end{align}
where $\tilde{f}_0$ is defined by 
$f_0(\varepsilon_n,{\bf p}_{\rm F},{\bf r})
=\tilde{f}_0(\varepsilon_n,{\bf p}_{\rm F},{\bf r}){\rm e}^{i\varphi({\bf r})}$, 
${\bf v}_{\rm s}$ is the superfluid velocity
${\bf v}_{\rm s}\equiv(\hbar/2m)({\bm\nabla\varphi-2e{\bf A}/\hbar})$, 
and $\varphi$ is the phase of the pair potential defined by 
$\Delta({\bf r})=|\Delta({\bf r})|{\rm e}^{i\varphi({\bf r})}$. 
At low temperatures and weak fields, the slope of the gap amplitude increases due to the core shrinkage
known as the Kramer-Pesch effect \cite{KP74}, which gives rise to a large correction to the charge density at the core. Thus, the core charge is enhanced more drastically  at lower temperatures and weak fields. 
On the other hand, the core charge due to the Lorentz force has a peak formed by the competition between the increasing magnetic field 
and the decreasing pair potential \cite{KUK16}. 
Thus, the PPG terms, which is dominant for charging at weak fields but has no explicit magnetic-field dependence,
is overwhelmed eventually by the Lorentz force as the field is increased.
We also find that 
the vortex-core charge at $T=0.5T_{\rm c}$ is much smaller than that at $T=0.2T_{\rm c}$ 
in all magnetic fields. 
Moreover, since the upper critical field becomes lower as the temperature is increased, 
the region where the PPG terms become dominant also becomes relatively wider.

We finally present an order-of-magnitude estimate for 
the accumulated charge $Q$ in the core region of radius $0.2 \xi_0$ from the core center. 
Figure \ref{fig-rho0}(a) gives the peak value $\rho({\bf 0})\!\simeq\! 70\rho_0\! =\!70(\Delta_0\epsilon_0/|e|^2\xi_0^2)|e|$ 
of the core-charge density  at  $T\!=\!0.2T_{\rm c}$ and $\bar{B}\!=\!0.7315B_{{\rm c}2}$.
Choosing $\Delta_0\!\approx\! 30 \ {\rm meV}$ as appropriate for YBa$_2$Cu$_3$O$_{7-x}$ (YBCO), 
we can estimate the peak value of the vortex-core charge $Q$ in a region of radius $0.2\xi_0$ from the core center and
length $d\,{\rm nm}$ along the $c$ axis as $Q \!\sim\! 0.01|e|d$. 
This value
is two orders of magnitude larger than 
the charge reported in Ref.\ \cite{KUK16}, owing to 
the different calculation method as explained around
Eqs.\ (\ref{g1_Lorentz1}) and (\ref{g1_Lorentz2}). 
It should be pointed out finally that, 
although our estimate on the vortex-core charge in YBCO is based on an $s$-wave model, 
one may expect that the energy-gap anisotropy does not much affect the magnitude of the accumulated charge in the core; see also Ref.\ \onlinecite{KUK17} on this issue. This issue certainly needs to be studied in more detail in the future by solving the AQC equations for anisotropic pairings.

\section{Conclusion \label{sec.IV}}
We developed a numerical method for studying charging 
in the vortex lattices of type-II superconductors based on the AQC equations 
with the Lorentz force and PPG terms. 
Using it, we calculated magnetic-field dependence of the vortex-core charge 
and spatial profile of the charge density in the vortex lattice 
of $s$-wave superconductors with a cylindrical Fermi surface 
at $T=0.2T_{\rm c}$ and $0.5T_{\rm c}$. 
We showed that, at both temperatures, 
the vortex-core charge is dominated by the PPG terms near the lower critical field
and by the Lorentz force near the upper critical field. 
Since the the upper critical field gets larger as the temperature is lowered, 
the dominant region of the core charge due to the Lorentz force at $T=0.2T_{\rm c}$ 
becomes wider than that at $T=0.5T_{\rm c}$. 
We also showed that 
the sign of accumulated charge due to the Lorentz force
can be understood in terms of the force picture,   
when the current density, the magnetic field, and the signs of the carriers are given. 
On the other hand, whether or not such a force picture is possible for the charging due to the PPG terms remains an open question.

There are other interesting problems 
on the physics of vortex lattice systems that may be tackled by the AQC equations. 
For example, the present method can be used to study the flux-flow Hall effect 
in the vortex lattice state by combining them with the AC response theory 
based on the standard Eilenberger equations 
\cite{Eschrig99,Eschrig09}. They can also be applied to vortex lattices 
in superfluid $^3$He \cite{Salomaa83,Thuneberg,Kita01-2,Kita02,Regan20}. 
Since spin currents may flow around the vortices in superfluid $^3$He, it may also be worthwhile to calculate the rotation-speed dependence of spin accumulation at the core and the spin flows due to the spin-accumulated vortices 
moving along transport mass currents.

Kumagai {\it et al}.\ studied the vortex-core charge in cuprate superconductors by the NMR/NQR measurements \cite{Kumagai01}. They estimated the accumulated charge by the local electric-field gradient signaled by changes in the nuclear quadrupole resonance frequency. To the best of our knowledge,
no direct observations of the vortex-core charge have been performed yet. 
We hope that the present study will stimulate them, such as those using the atomic force microscopy technique.

\begin{acknowledgments}
We thank T. Uchihashi, J. Goryo, R. C. Regan, A. Kirikoshi, 
and E. S. Joshua for useful discussions and comments. 
H.U. is supported in part by JSPS KAKENHI Grant No. 15H05885 (J-Physics). 
The computation in this work was carried out using the facilities of Supercomputer Center, 
Institute for Solid State Physics, the University of Tokyo.
\end{acknowledgments}

\nocite{*}

\bibliography{PPG_charge_Abrikosov}

\end{document}